# IMPORTANCE OF REALISTIC MOBILITY MODELS FOR VANET NETWORK SIMULATION

Bahidja Boukenadil[1]

[1]Department of Telecommunication, Tlemcen University, Tlemcen, Algeria

*ABSTRACT*

*In the performance evaluation of a protocol for a vehicular ad hoc network, the protocol should be tested under a realistic conditions including, representative data traffic models, and realistic movements of the mobile nodes which are the vehicles (i.e., a mobility model). This work is a comparative study between two mobility models that are used in the simulations of vehicular networks, i.e., MOVE (MObility model generator for VEhicular networks) and CityMob, a mobility pattern generator for VANET. We describe several mobility models for VANET simulations.*

*In this paper we aim to show that the mobility models can significantly affect the simulation results in VANET networks. The results presented in this article prove the importance of choosing a suitable real world scenario for performances studies of routing protocols in this kind of network.*

*KEYWORDS*

*VANET; Mobility Model; Simulations; Real World; MOVE; SUMO; CityMob; NS-2 etc.*

## 1. INTRODUCTION

The mobility is the principal constraint met in the vehicular networks, therefore the performance investigation of VANET network routing require the exact prediction of mobility of nodes forming the network, and this is realized by the good choice of the mobility model.

Generally, the traces and the synthetic models [1] are two kinds of mobility patterns used for the simulation of networks .for simulating the complex and large scale networks that require more real world scenarios for a real modeling, the traces are used. But in several situations, these traces are created with difficulty, particularly for the networks of high mobility like the vehicular networks. In this case, the mobility of nodes (i.e., the vehicles nodes) are modeled with the synthetic patterns, these models try to really present the behavior of these environments of vehicles.

A variety of synthetic mobility models for vehicular networks are discussed in Literature, some of them are presented in Section II. Section III illustrate that a mobility model has a large effect on the performance evaluation in simulation of VANET network. Finally, Section IV presents some concluding remarks.

## 2. MOBILITY MODEL FOR VEHICULAR NETWORKS

Many models of mobility can be used to generate the movement of the vehicles instead of the traces which are difficult to create in the VANET environments .

However, the behavior of these networks is very complex; the drivers must react to the high change of the road conditions like the congestion, traffic jam, works in the roads. The state of the roads in turn depends on the behaviors of the drivers. Thus, the choice of the mobility model affects the relevance and viability of the obtained results.

In general, vehicular traffic simulators or mobility models can be classified in two great kinds, the microscopic and macroscopic simulators. The microscopic simulator is interested in the movement of each vehicle taking part in the road traffic. In contrast the macroscopic simulators are interested in all the vehicles forming the networks; it compute road capacity and the distribution of the traffic in the road net, by determining a certain parameters like traffic density (number of vehicles per km per lane) or traffic flow (number of vehicles per hour crossing some points of interest, usually an areas with a great density of nodes) .

The literature present a large variety of mobility models for VANET networks.

The work of Saha and Johnson [2] model vehicular traffic in real road topologies of the maps of USA available from TIGER (Topologically Integrated Geographic Encoding) [3] database. The mobility of nodes is random.Vehicles compute the shortest path to get their destination over the graph by weighting the cost of displacement on each road on its speed limit and the traffic jam.

Huang et al. [4] studied taxi behavior. In this work, a Manhattan style grid with a uniform block size is modeled. The streets of simulation area are two-way, with one lane in each direction. These lanes constrain vehicle movements. A set of parameters characterize the vehicle behavior which are preferred velocity ,maximum acceleration and deceleration, a velocity variation associated with the preferred speed at steady state and a list of preferred destinations, i.e., the taxi stands. . The taxis assigne randomly one of three preferred speeds.

In the work of Choffnes et al. [5] , an integrated mobility and traffic model is conceived, named STRAW.For modeling real traffic conditions ,this model incorporates a simple car-following model with traffic control. The road map of this pattern is constructed by street plans with one lane in each direction .The vehicles move in these lanes according to a random street placement model.Each vehicle enters the map is placed behind the existing one.

In [6] Haerri et al. proposed a vehicular mobility simulator for VANET networks, called VanetMobiSim [7] .The speed of vehicles for this model is determined by the use of the Intelligent Driver Model (IDM).

Three different models were presented in the work of Mahajan et al. [8]: Stop Sign Model (SSM), Traffic Light Model (TLM) and Probabilistic Traffic Sign Model (PTSM).In this models All roads are two-way, with one lane in each direction for the SSM and PTSM, whereas TLM model streets with multiple lanes. An algorithm is employed to reproduce stop signs for each model.

Martinez et al [9] present CityMob [10] .This mobility model generate different mobility patterns in VANETs, Simple Model (SM), Manhattan Model (MM) and Downtown Model (DM).Each model has its own characteristics,For the SM,the plan of the road is very simple, neither semaphores nor direction changes.In constract, MM and DM simulates semaphores at crossing and random positions in zones with different vehicle density. CityMob allows to simulate damaged cars and tries to prevent congestions. Figure 1 present the interface of this model.

Karnadi et al. [12, 13, 14] develop a tool MOVE (MObility model generator for VEhicular networks) [11] .This model provide facility for the users to generate real world scénarios in VANET environments. This model contains two editors, one for road map and the second for vehicle movement .The user can manually and automatically create the road topology or import it from existing real world maps such as Google maps. For vehicle movement, the pattern can be manually created, generated automatically or specified based on a bus time table to simulate the movements of public transportations. Figure2 shows the interface of this model.

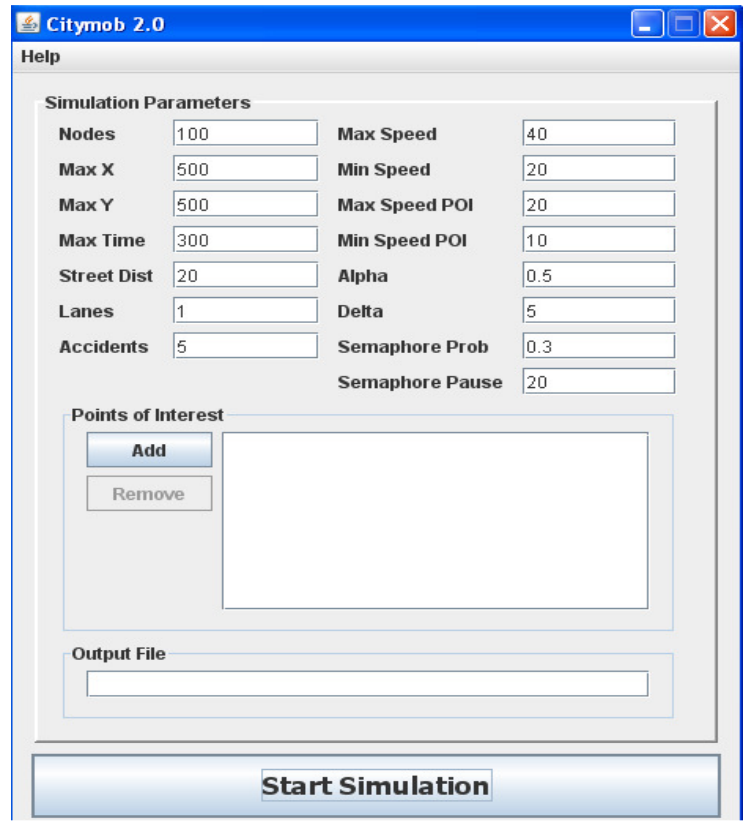

Figure 1. Interface of CityMob

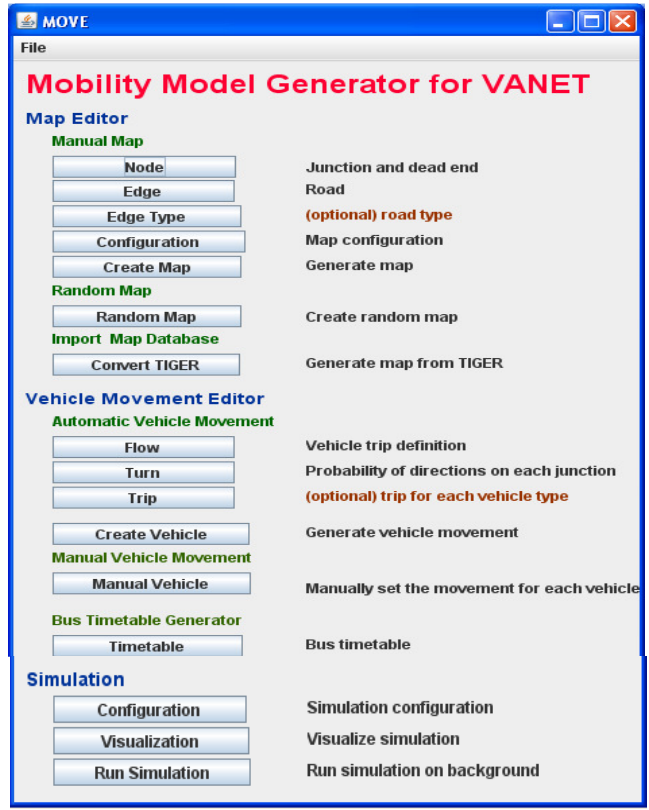

Figure 2: Mobility Model Generator

## 3. INFLUENCE OF MOBILITY MODEL IN VANET SIMULATIONS

The simulation results are greatly affected by the mobility model selected .The results presented in this section illustrate the importance of choosing a suitable mobility model for performances evaluation of routing protocols in VANET network.

We use ns-2 [34] to compare the performance of the CityMob Model and the MOVE tool via a simulation.The routing of packets is accomplished with the Ad hoc On Demand Distance Vector) (AODV) [20]. We have chosen the parameters for both mobility models in a manner to simulate the same path as possible.

AODV (Ad hoc On Demand Distance Vector) [1] is a reactive protocol that does route discovery when needed by a node. The route discovery process is initiated only when a source node has data traffic to send to a destination node, that makes AODV a truly On- Demand routing protocol.

In our simulations, we chose AODV since it behaves well in several of the performance evaluations of the routing protocols (e.g. [21, 22, 23]).

178

International Journal of Computer Networks & Communications (IJCNC) Vol.6, No.5, September 2014

The ns-2 code used in our simulations of AODV was obtained from [24].
Each simulation run lasted for 300 seconds with a uniform block size of 500 x 500 meters; the maximum speed of vehicles is of 40 m/s.

The number of vehicles nodes from 15 to 100. We chose the traffic sources at a constant rate CBR (Constant Bit Rate). Traffic between nodes is generated using a traffic generator which is characterized by the following parameters: Data packets Size is 512 bytes, Interval between packets is 0.25s , maximum of packets transmitted 1000. All nodes use IEEE 802.11 MAC operating at 2Mbps. The propagation model employed in the simulation is TwoRayGround reflextion.

In our comparison of the two mobility models, we consider the following performance metrics obtained from the AODV protocol: throughput, end-to-end delay and protocol overhead.

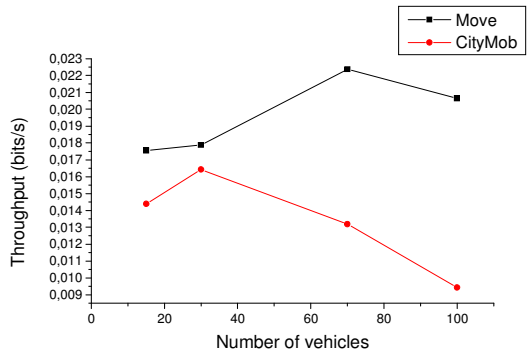

Figure 3 : Throughput vs. number of vehicles.

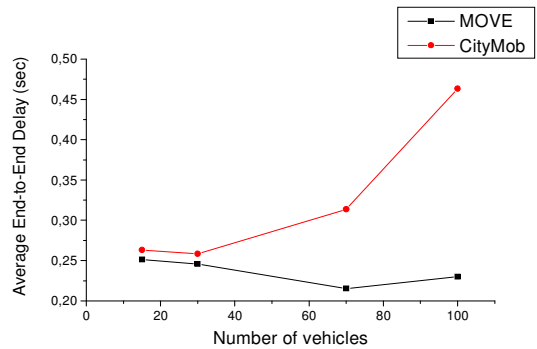

Figure 4 : End-to-end delay vs. number of vehicles.

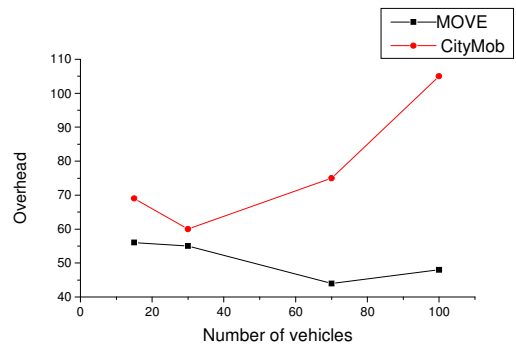

Figure 5 : overhead vs. number of vehicles.

Figures 3, 4 and 5 illustrate the performance (i.e., Throughput ratio, end-to-end delay and overhead) of AODV with the two mobility models chosen.The Throughput (in figure 3) of AODV when using MOVE mobility model is higher and more stable than when using CityMob model.The trace 4 shows that AODV causes a low stable delay with MOVE because the roads are more defined compared to CityMob.

Figures 5 illustrate the overhead AODV required with each of the chosen mobility models. The vehicles moving with CityMob have a higher overhead, as a result this model requires a higher amount of overhead compared to MOVE.These results confirm the suitability of MOVE tool for simulating VANET.

## 4.CONCLUSIONS

In this paper, we compared the performance of two mobility models for VANET Simulation i.e. MOVE (MObility model generator for VEhicular networks) and CityMob (City Mobility).Simulation analysis using realistic mobility model for VANET environment show that the performance of the protocol is greatly affected by the mobility model. The performance of an ad hoc network protocol can vary significantly with different mobility models then, the choice of mobility model in simulating VANET is very important.The mobility models for VANET should be most closely match the expected real-world scenario. In fact, the realistic scenarios can aid the development of the routing protocols significantly.

As future work we plan to compare other mobility models discussed above. Results obtained from these studies would certainly facilitate in meeting the challenges associated with future development and evaluation of suitable routing protocols in vehicular networks.